\documentclass[twocolumn,showkeys,showpacs,preprintnumbers,amsmath,amssymb,superscriptaddress]{revtex4}

\usepackage{epsfig}
\usepackage{graphicx}% Include figure files
\usepackage{dcolumn}% Align table columns on decimal point
\usepackage{bm}% bold math
\usepackage{subfigure}
\usepackage{color,soul}
\graphicspath{{./}{./Fig/}}

\begin{document}

% \preprint{preprint: BUS15}
% \preprint{APS/123-QED}

\title{Chimera states in time-varying complex networks}

\author{Arturo Buscarino}
\author{Mattia Frasca}
\email[corresponding author: ]{mfrasca@diees.unict.it}
\author{Lucia Valentina Gambuzza}
\affiliation{DIEEI, Universit\` a degli Studi di Catania, Catania, Italy}
\author{Philipp H{\"o}vel}
\affiliation{Institut f{\"u}r Theoretische Physik, Technische Universit{\"a}t Berlin, Hardenbergstra\ss{}e 36, 10623 Berlin, Germany}
\affiliation{Bernstein Center for Computational Neuroscience, Humboldt-Universit{\"a}t zu Berlin, Philippstra{\ss}e 13,
10115 Berlin, Germany}

\date{\today}

\begin{abstract}
Chimera states have been recently found in a variety of different coupling schemes and geometries. In most cases, the underlying coupling structure is considered to be static, while many realistic systems display significant temporal changes in the pattern of connectivity. In this work, we investigate a time-varying network made of two coupled populations of Kuramoto oscillators, where the links between the two groups are considered to vary over time. As a main result, we find that the network may support stable, breathing and alternating chimera states. We also find that, when the rate of connectivity changes is fast, compared to the oscillator dynamics, the network may be described by a low-dimensional system of equations. Unlike in the static heterogeneous case, the onset of alternating chimera states is due to the presence of fluctuations, which may be induced either by the finite size of the network or by large switching times.
\end{abstract}

% Synchronization, nonlinear dynamics, 05.45.Xt
% Complex systems, 89.75.-k
% nonlinear dynamics, 05.45.-a
% Control of chaos, applications of chaos, 05.45.Gg


\maketitle

\section{Introduction}

The first evidence of chimera states dates back to 2002 \cite{kuramoto02}, when Kuramoto and Battogtokh, studying a system of identical phase oscillators coupled in a non-local way, discovered the onset of a counterintuitive behavior: the oscillators split into two coexisting subpopulations, one coherent and one incoherent. Since that first report, the phenomenon attracted a lot of interest leading to the discovery of chimera states in a variety of systems (phase oscillators \cite{kuramoto02,abrams04,shima04,abrams08,martens10,motter10}, neurons \cite{singh13,CSnostro}, chemical units \cite{tinsley12}, chaotic units \cite{CSchaoticMap,CSchaoticUnits}). While chimera states were initially observed only in systems with non-local coupling (one-dimensional rings \cite{abrams04,kuramoto02}, two-dimensional systems \cite{shima04,martens10,motter10}) and for pure phase dynamics, the results of recent works pointed out the appearance of chimera states also in systems with global coupling or with not negligible amplitude dynamics \cite{singh13,sethia13,schmidt14,scholl14,sethia14,OME13}.

A structure particularly relevant for our study is the one formed by two coupled populations where each oscillator is equally coupled to all the others in its group, and less strongly to those in the other group \cite{abrams08}. Despite the symmetry of the coupling structure, an asymmetric behavior -- with one population displaying synchronized oscillations and the other exhibiting incoherence -- emerges in this network. The incoherent population may either show a constant level of desynchronization (\emph{stable chimera}) or an oscillating one (\emph{breathing chimera}). Notably, when the intrinsic frequencies of oscillators are not homogeneous, an \emph{alternating chimera}, where the two populations alternate in the level of synchrony, is observed \cite{Laing12}. This observation may be linked to unihemispheric sleep, where sleep alternates between the two hemispheres with one half of the brain awake with desynchronized neuronal activity and the other sleeping and synchronized \cite{Rattenborg,Mathews,Lyamin}. Alternating chimera states have been also found in coupled populations of forced oscillators \cite{Ma10}, in time-delayed systems \cite{SHE10} and in isotropic oscillatory media with nonlinear uniform global coupling \cite{Haugland14}. The onset of stable and breathing chimera states is not limited to two populations, but is found also in systems formed by more than two coupled populations \cite{martenschaos,martenspre}.

In recent works, the concept of chimera states has been generalized to include other types of symmetry breaking solutions and new terms have been coined: \emph{amplitude-mediated chimera} displaying temporal variations of the amplitude in the incoherent population \cite{sethia13}; \emph{amplitude chimera}, that is, a chimera behavior of the oscillator amplitude rather than its phase \cite{CSchaoticMap,CSchaoticUnits}; \emph{chimera death} \cite{scholl14}, characterized by coexistence of spatially coherent and incoherent oscillation death; \emph{chimera states with quiescent and synchronous domains} (QSCS), where synchronization coexists with spatially patterned oscillation death \cite{singh13,CSnostro}. In parallel to theoretical investigations, experimental studies have demonstrated the existence of chimera states in real systems. In \cite{hagerstrom12} chimera states have been revealed in a coupled map lattice made of a liquid-crystal spatial light modulator; in \cite{tinsley12} a system of coupled Belousov-Zhabotinsky oscillators has shown chimera behaviors such as phase-cluster states; in \cite{martens13} chimera states have been observed in a set of metronomes placed on two weakly coupled swings. An experimental evidence of QSCS is reported in \cite{CSnostro} for a system of electronic circuits with neuron-like spiking dynamics.

Most of the works on chimera states assume that the connection structure is static. However, in many systems (for example, communication, ecological, social, contact networks) links are not always active and the connectivity between units changes during time with a rate ranging from slow to fast \cite{holme12}. The dynamics of the systems interacting through a network can be significantly affected by the link activity. For this reason, the pattern of link activation is explicitly taken into account as an element of the system in the study of time-varying or temporal networks \cite{holme12}. The dynamics of time-varying networks is characterized by the presence of two time scales (those of the dynamical process and that of the link activation) and by the rule (which can be either deterministic or stochastic) defining the connectivity changes in time. In several works \cite{Porfiir06,Belykh04,Porfiri12}, to account for sporadic intermittent interactions, time-dependent connections are introduced by switching on or off, at a fixed frequency, a subset or the whole set of the edges of a network. For this setting, an analytical approach for global synchronization is derived in the limit of fast switching. In this paper, we use this framework to study the onset of chimera states in a time-varying network. In particular, we consider a system made of two coupled populations with strong, time-independent links within each group and less strong interconnections between them modeled by time-dependent edges. We found that the system may exhibit stable, breathing and alternating chimera states. Alternating chimera states are found when the fluctuations due to the stochastic switching of the connections are not negligible.

The rest of this paper is organized as follows: Section~\ref{sec:Model} introduces the model equations and network structure and presents a bifurcation diagram. Section~\ref{sec:reduced} discusses a low-dimensional set of reduced equations to illustrate the mechanism of switching. Section~\ref{sec:alternatingchimera} addresses questions related to the size of the populations. Finally, we summarize the results in Sec.~\ref{sec:conclusion}.

\section{A system of two coupled populations with time-varying interactions}
\label{sec:Model}
We consider a pair of oscillator populations where the coupling between groups changes as a function of time. Each population $\sigma$ (with $\sigma=1,2$) consists of $N_\sigma$ identical phase oscillators. Within each population the oscillators are globally coupled with links fixed in time and of weight $\mu$, while the coupling between the two populations is unitary and time-varying. The inter-population links are randomly switched on or off at fixed equally spaced time intervals of length $\tau$. During each time interval, every possible connection between two nodes in different groups is turned on, with probability $p$, independently of the other links, and independently of whether or not it has been turned on during the previous time interval. This leads to an inter-population connectivity which is a time-varying matrix given by a random sequence of Erd\H{o}s-R\'enyi graphs with average in-degree $p N_\sigma$. The system of two interacting populations is described by:
\begin{equation}
\label{eq:modelcoupledpopulations}
\frac{d}{dt}\theta_i^\sigma=\omega+\sum_{\sigma^{'}=1}^2 \frac{1}{N_{\sigma'}} \sum_{j=1}^{N_{\sigma'}} K^{\sigma
\sigma'}_{ij}(t)
\sin (\theta_j^{\sigma'}-\theta_i^\sigma-\alpha),
\end{equation}
where $\theta_i^\sigma$ is the phase of oscillator $i$ in population $\sigma$, $\omega$ is the intrinsic
frequency (equal for all the oscillators, fixed at $\omega=1$), $\alpha$ is the phase lag, and
$K^{11}_{ij}(t)=K^{22}_{ij}(t)=\mu > 0$ $\forall t$. $\mathbf {K}^{12}(t)=(\mathbf{K}^{21}(t))^T$ are stochastic
matrices whose elements are defined as $K^{12}_{ij}(t)=K^{21}_{ji}(t)=s_{ij}(q)$ for $(q-1)\tau<t<q\tau$ with:
\begin{equation}
s_{ij}(q)=\left\{ \begin{array}{ll}
1 & \textrm{with probability}~p \\ 0 & \textrm{with probability}~1-p
\end{array} \right.
\end{equation}
where $q\in \mathcal{N}^+$ defines the number of switching intervals, each of length $\tau$.

To monitor coherence in each population, two separate Kuramoto order parameters are considered:
\begin{equation}\label{eq:order_param}
r_\sigma(t)=\left|\left\langle e^{\iota \theta_i(t)}\right\rangle_\sigma\right|
\end{equation}
with $\sigma=1,2$ and $\iota=\sqrt{-1}$. $\langle \cdot \rangle_\sigma$ denotes the average over all elements in population $\sigma$.

\begin{figure}
\centering
\includegraphics[width=0.8\linewidth]{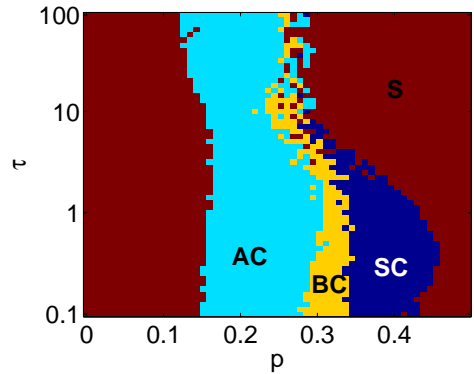}
\caption{(Color online) Bifurcation map with respect to the switching probability $p$ and to the length of the switching interval $\tau$. The population size is $N_1=N_2=100$, the coupling strength within each group is fixed to $\mu=0.6$, the oscillators frequency $\omega=1$, and the phase lag $\alpha=1.5$. The regions are labelled according to the behavior observed: S synchronization; SC stable chimera; BC breathing chimera; AC alternating chimera. \label{fig:bifDiagram}}
\end{figure}

\begin{figure*}
\centering
\subfigure[]{\includegraphics[width=0.32\linewidth]{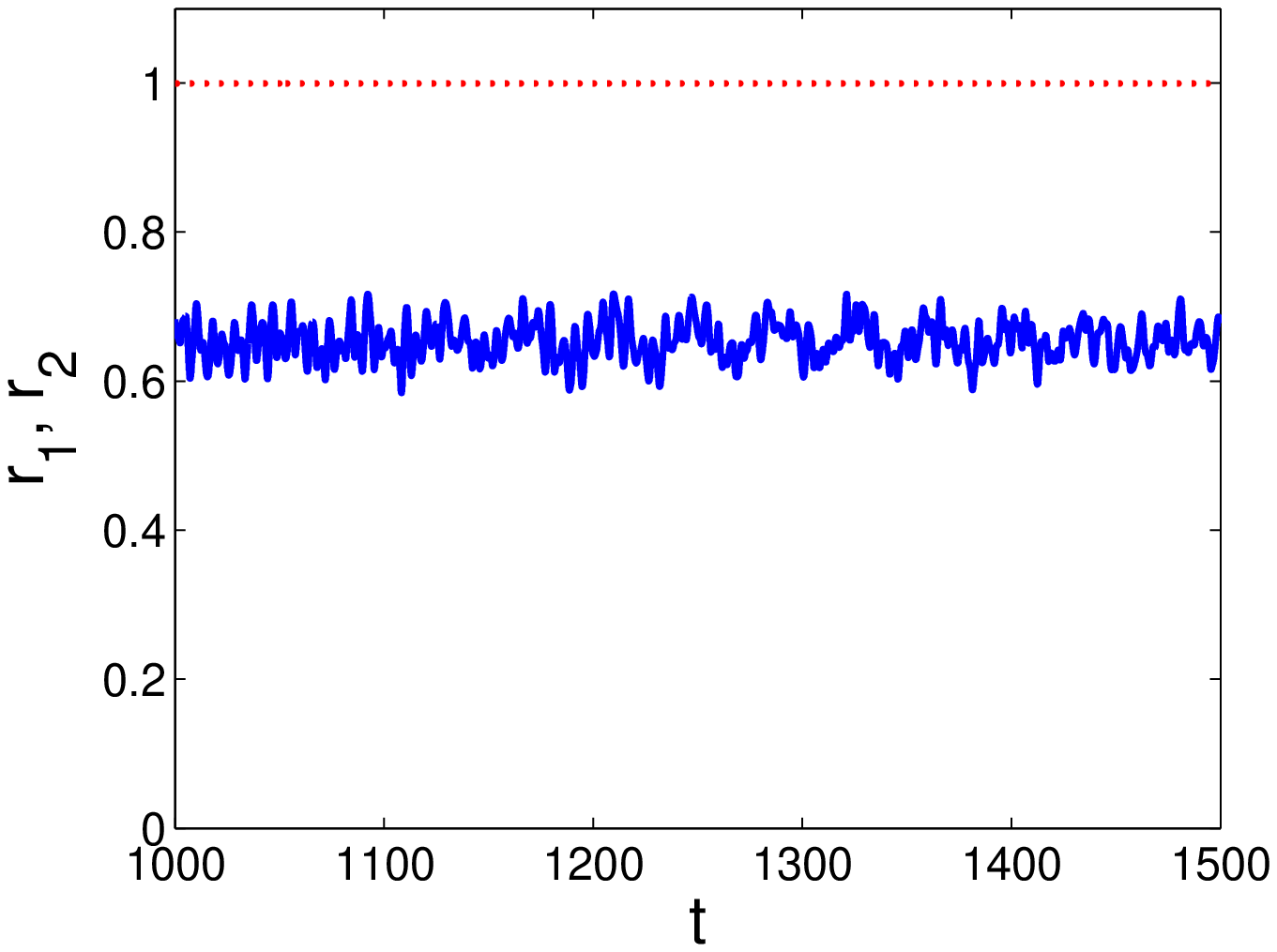}\label{fig:CSClassicoB}}
\subfigure[]{\includegraphics[width=0.32\linewidth]{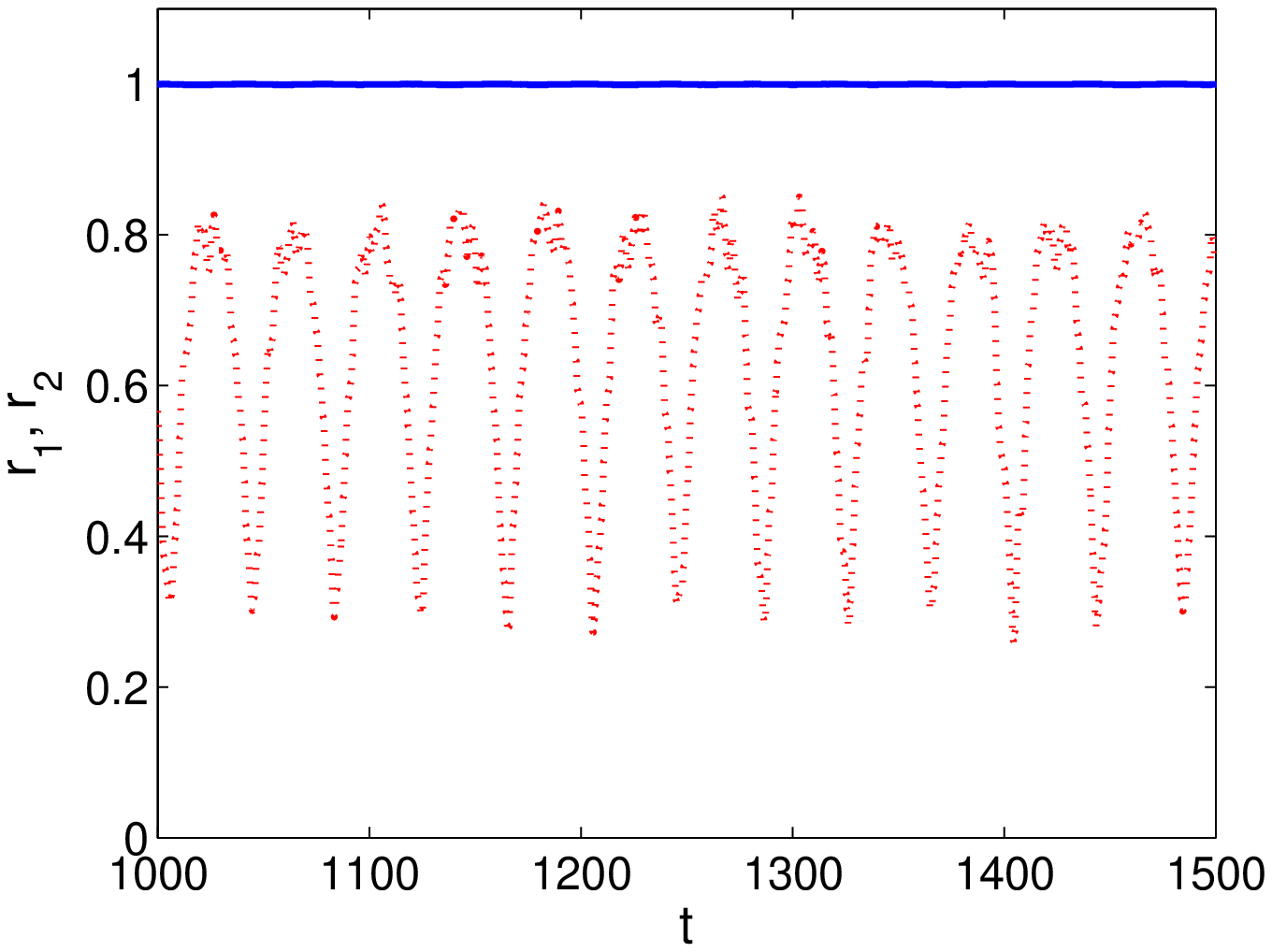}\label{fig:CSBreathingB}}
\subfigure[]{\includegraphics[width=0.32\linewidth]{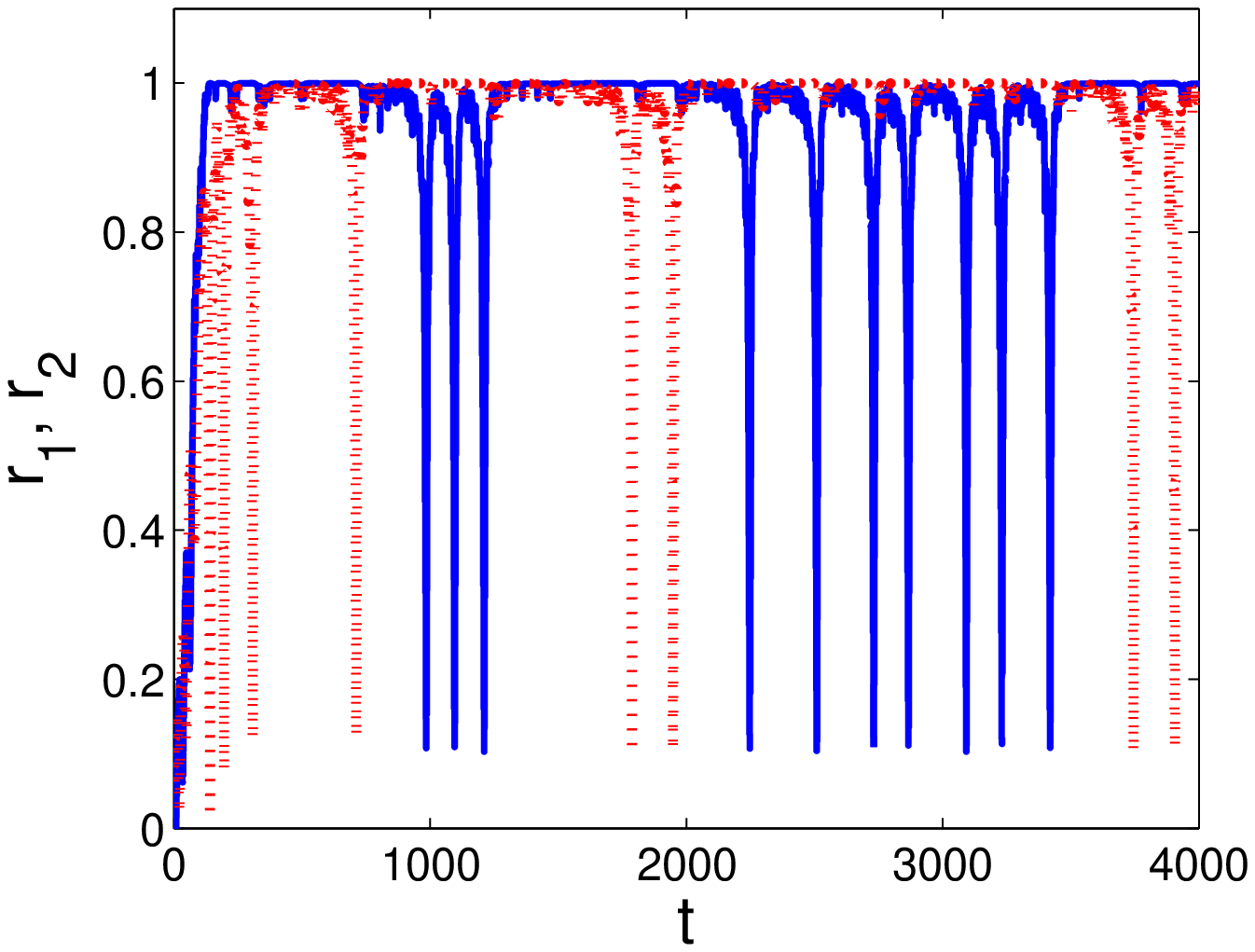}\label{fig:CSBlinkingB}}
\caption{(Color online) Time series of the two order parameters $r_1(t)$ (blue solid) and $r_2(t)$ (red dotted) for $\tau=0.1$ and different values of the switching probability $p$: (a) $p=0.38$; (b) $p=0.33$; (c) $p=0.25$. Other parameters as in Fig.~\ref{fig:bifDiagram}. \label{fig:CSall}}
\end{figure*}

To illustrate the effect of the switching of the inter-population links, we discuss the behavior of a network with $N_1=N_2=N=100$ oscillators by varying the values of the parameters ruling the switching, that is, the probability $p$ and the length $\tau$ of the time intervals. The bifurcation diagram, shown in Fig.~\ref{fig:bifDiagram}, reveals the onset of different types of chimera states in a large region of the parameter space $(p,\tau)$. The region labelled as S is characterized by synchronization of both populations, $r_1=r_2 \simeq 1$. All the other regions indicate coexistence of synchronization with a chimera state. These are illustrated in Fig.~\ref{fig:CSall}, where the evolution of the two Kuramoto order parameters is reported for selected values of $p$ with $\tau$ fixed to $\tau=0.1$. Stable chimeras are found in the region SC (cf. Fig.~\ref{fig:bifDiagram}) and are characterized by one coherent population, showing synchronized oscillations and an order parameter close to one (population 2 in Fig.~\ref{fig:CSall}(a)) and one desynchronized (population 1). The phase coherence for the desynchronized population remains approximately constant. For breathing chimeras (region BC in Fig.~\ref{fig:bifDiagram}), instead, the phase coherence of the desynchronized population is not constant, but pulsates (Fig.~\ref{fig:CSall}(b)). Alternating chimera states appear in the region AC. These chimeras are characterized by alternating synchrony between the two populations (Fig.~\ref{fig:CSall}(c)). While one population is nearly synchronized, the other displays a pulsating order parameter; the oscillators in the desynchronized population may then gain synchrony at the expense of the oscillators in the other population which lose  synchrony. The behavior is found to alternate with either regular or irregular periods as a function of the value of $p$ and $\tau$.

We note that, when the two populations are coupled with time-varying links, stable, breathing and alternating chimeras are all observed for identical oscillators. Moreover, when the pattern of connectivity is fixed in time, if the intrinsic frequencies are homogenous, only stable and breathing chimeras appear \cite{abrams08}, while the onset of alternating chimeras requires heterogeneity of the oscillators \cite{Laing12}.

\section{Reduced equations}
\label{sec:reduced}
In the thermodynamic limit of infinite system size, $N\rightarrow \infty$, many high-dimensional systems show low-dimensional dynamics. These systems may be reduced to a small set of ordinary differential equations for the study of the macroscopic evolution. This has been recently demonstrated for a system of globally coupled Kuramoto oscillators, which is reduced to a single first-order ordinary differential equation \cite{ott2008}, and then generalized to assortative networks \cite{restrepo14}. In this Section, we write down a low-dimensional model for Eqs.~(\ref{eq:modelcoupledpopulations}) and show that this is able to explain the occurrence of stable and breathing chimeras in our system. The mechanism underlying the onset of alternating chimera states will be discussed in Section~\ref{sec:alternatingchimera}.

We first introduce a non-switching system, obtained from Eqs.~(\ref{eq:modelcoupledpopulations}) by considering a time-averaged connectivity:
\begin{equation}
\label{eq:averagedsystem}
\frac{d}{dt}\theta_i^\sigma=\omega+\sum_{\sigma^{'}=1}^2 \frac{ \left\langle K^{\sigma \sigma'}_{ij}
\right\rangle}{N_{\sigma'}} \sum_{j=1}^{N_{\sigma'}} \sin (\theta_j^{\sigma'}-\theta_i^\sigma-\alpha)
\end{equation}
with
\begin{equation}
\left\langle K^{\sigma \sigma'} \right\rangle=\left\{ \begin{array}{ll}
\mu & \textrm{if}~\sigma=\sigma' \\ p & \textrm{if}~\sigma \neq \sigma'
\end{array} \right.
\end{equation}

Under the assumption that the switching period is small, that is, the changes of the network topology operate on a time scale faster than the node dynamics, it is to be expected that the behavior of the switching system in Eqs.~(\ref{eq:modelcoupledpopulations}) is close to that of the averaged system. This is also confirmed by several works investigating the effects of an increasing switching frequency \cite{Porfiir06,Belykh04,Porfiri12,Siam2013}.

By applying the Ott-Antonsen ansatz \cite{ott2008} to Eqs.~(\ref{eq:averagedsystem}), the dynamics of the averaged system is then described in terms of the oscillator density distribution $f^\sigma(\theta)$. Omitting a detailed derivation, one obtains the following set of reduced equations
\begin{subequations}
\label{eq:rho1_rho2_psi}
\begin{align}
\dot{\rho}_\sigma&= \frac{1-\rho_\sigma^2}{2} \sum_{\sigma'=1}^2 \left\langle K^{\sigma \sigma'} \right\rangle
\rho_{\sigma'} \sin (\phi_{\sigma'} - \phi_\sigma + \beta) \\
\dot{\phi}_\sigma&= \omega-\frac{1+\rho_\sigma^2}{2\rho_\sigma} \sum_{\sigma'=1}^2 \left\langle K^{\sigma \sigma'}
\right\rangle \rho_{\sigma'} \cos(\phi_{\sigma'} - \phi_ \sigma + \beta),
\end{align}
\end{subequations}
where we used $\beta=\pi/2-\alpha$. Defining the phase difference between the two populations $\psi=\phi_1-\phi_2$ yields the following equations
\begin{subequations}
\label{eq:rho1_rho2_psi2}
\begin{align}
\dot{\rho}_1=& \frac{1-\rho_1^2}{2}[\mu \rho_1 \cos \alpha + p \rho_2 \cos(-\psi-\alpha)]\\
\dot{\rho}_2=& \frac{1-\rho_2^2}{2}[\mu \rho_2 \cos \alpha + p \rho_1 \cos(\psi-\alpha)] \\
\dot{\psi}=& -\frac{1+\rho_1^2}{2}\left[\mu \sin \alpha + p \frac{\rho_2}{\rho_1} \sin(\psi+\alpha)\right] \nonumber\\
& + \frac{1+\rho_2^2}{2}\left[\mu \sin \alpha + p \frac{\rho_1}{\rho_2} \sin(-\psi+\alpha)\right].
\end{align}
\end{subequations}

\begin{figure}
\centering
\includegraphics[width=\linewidth]{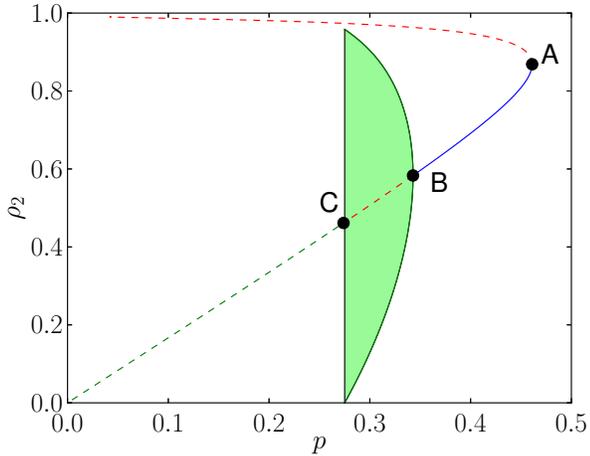}
\caption{(Color online) Bifurcation diagram of system (\ref{eq:rho1_rho2_psi2}) with respect to parameter $p$. Solid (dashed) curves indicate stable (unstable) fixed points. Green shading refers to a stable limit cycle. Point A, B, and C mark the saddle-node, Hopf, and homoclinic bifurcation, respectively. Parameters: $\mu=0.6$ and $\alpha=1.5$.
\label{fig:bif}}
\end{figure}

System~(\ref{eq:rho1_rho2_psi2}) is studied with respect to the parameter $p \in [0,0.5]$. Beyond the trivial equilibrium point $(1,1,0)$ which represents global synchronization of the network, the system has two further equilibria and an additional invariant limit cycle depending on $p$ as discussed below. Due to symmetry with respect to coordinates change $(\rho_1,\rho_2,\psi)\rightarrow(\rho_2,\rho_1,-\psi)$, it suffices to study only the equilibria on one of the planes $\rho_1=1$ or $\rho_2=1$.

%\begin{figure}
%\centering
%% \includegraphics[width=\linewidth]{bif.eps}
%\includegraphics[width=\linewidth]{fig_lc_period.eps}
%% \includegraphics[width=\linewidth]{fig_bd.png}
%\caption{(Color online) Period of the limit cycle present in system~(\ref{eq:rho1_rho2_psi2}) in dependence on $p$.
%Point B marks the Hopf bifurcation. Parameters: $\mu=0.6$ and $\alpha=1.5$.
%\label{fig:lc_period}}
%\end{figure}

\begin{figure}
\centering
\includegraphics[width=\linewidth]{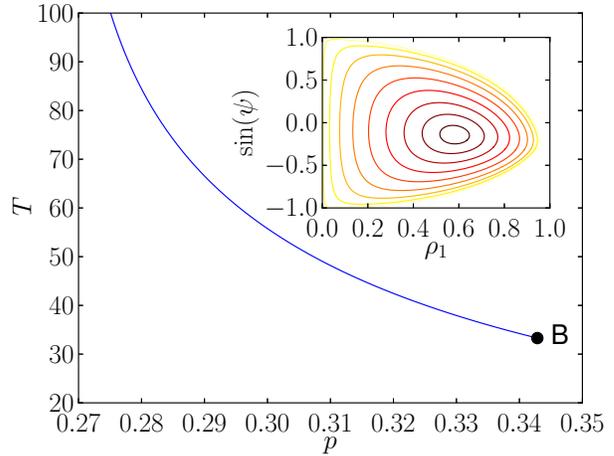}
\caption{(Color online) Period of the limit cycle present in system~(\ref{eq:rho1_rho2_psi2}) in dependence on $p$. Point B marks the Hopf bifurcation. The inset shows the limit cycle in the $(\rho_1,\sin(\psi))$-plane. The color gradient corresponds to the $p$-values (dark: larger, bright: smaller). Parameters: $\mu=0.6$ and $\alpha=1.5$.
\label{fig:lc_period}}
\end{figure}

Figure~\ref{fig:bif} depicts the bifurcation diagram of system~(\ref{eq:rho1_rho2_psi2}). Starting from $p=0.5$ and decreasing this parameter, we find a saddle-node bifurcation (point A in Fig.~\ref{fig:bif}), a Hopf bifurcation (point B in Fig.~\ref{fig:bif}), and a homoclinic bifurcation (point C in Fig.~\ref{fig:bif}). The different regions in the bifurcation diagram correspond to the onset of different types of chimera states. For $p \in [B, A]$ the system~(\ref{eq:rho1_rho2_psi2}) has three stable equilibrium points, which correspond to global synchronization or stable chimeras in one of the two populations. For $p \in [C, B]$ the system (\ref{eq:rho1_rho2_psi2}) has one stable equilibrium and -- due to the symmetry mentioned above -- two stable limit cycles, which give rise to a coexistence of global synchronization and breathing chimera states. The period of the this limit cycle increases for decreasing $p$ (see Fig.~\ref{fig:lc_period}) meaning that the period of the breathing chimera becomes longer as $p$ approaches the homoclinic bifurcation point. At the homoclinic bifurcation point C the limit cycles of system~(\ref{eq:rho1_rho2_psi2}) collide and annihilate in a homoclinic bifurcation so that for $p \in [0, C]$ only the trivial equilibrium $(1,1,0)$ persists.

We find that the reduced equations~(\ref{eq:rho1_rho2_psi2}) are effectively able to predict the behavior of the switching system for small switching periods and $p \in [C,0.5]$. Outside this region, alternating chimeras, not predicted by the reduced model, are found. As we will show in the next Section, the discrepancy between the prediction and the behavior observed in Fig.~\ref{fig:bifDiagram} is not due to a failure of the reduced model, but reflects a difference between a finite and an infinite network.

\begin{figure*}
\centering
\subfigure[]{\includegraphics[width=0.32\linewidth]{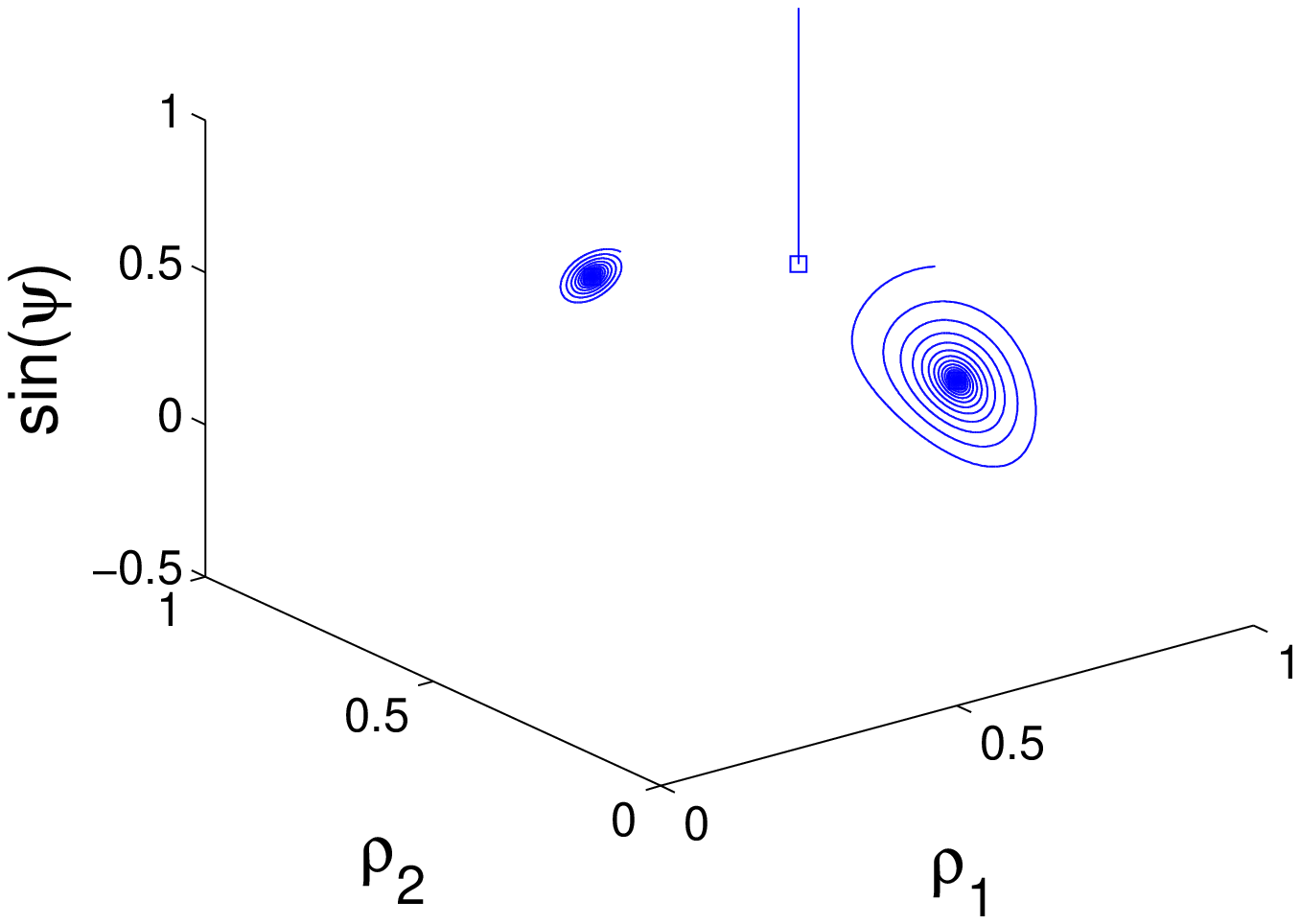}}
\subfigure[]{\includegraphics[width=0.32\linewidth]{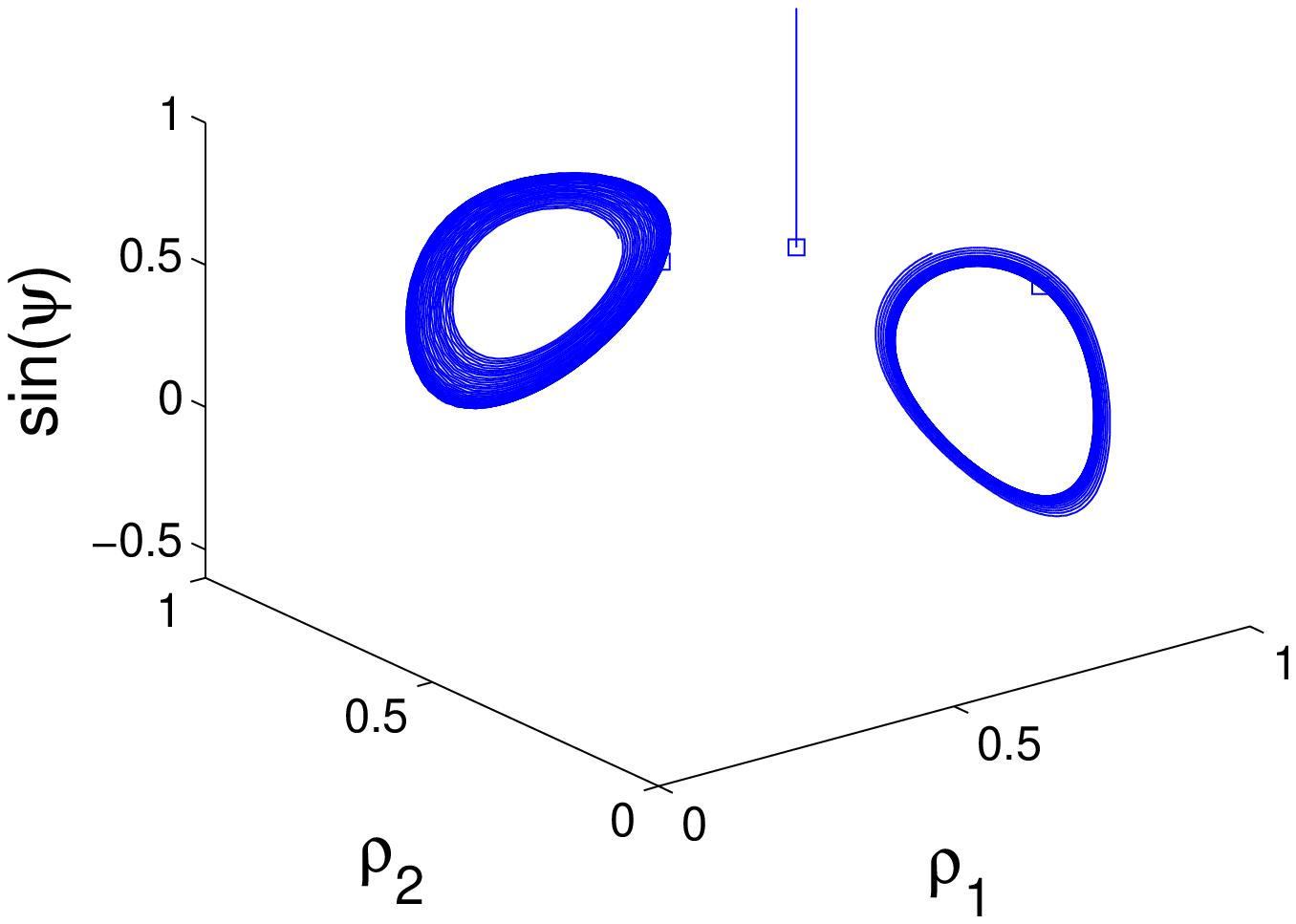}}
\subfigure[]{\includegraphics[width=0.32\linewidth]{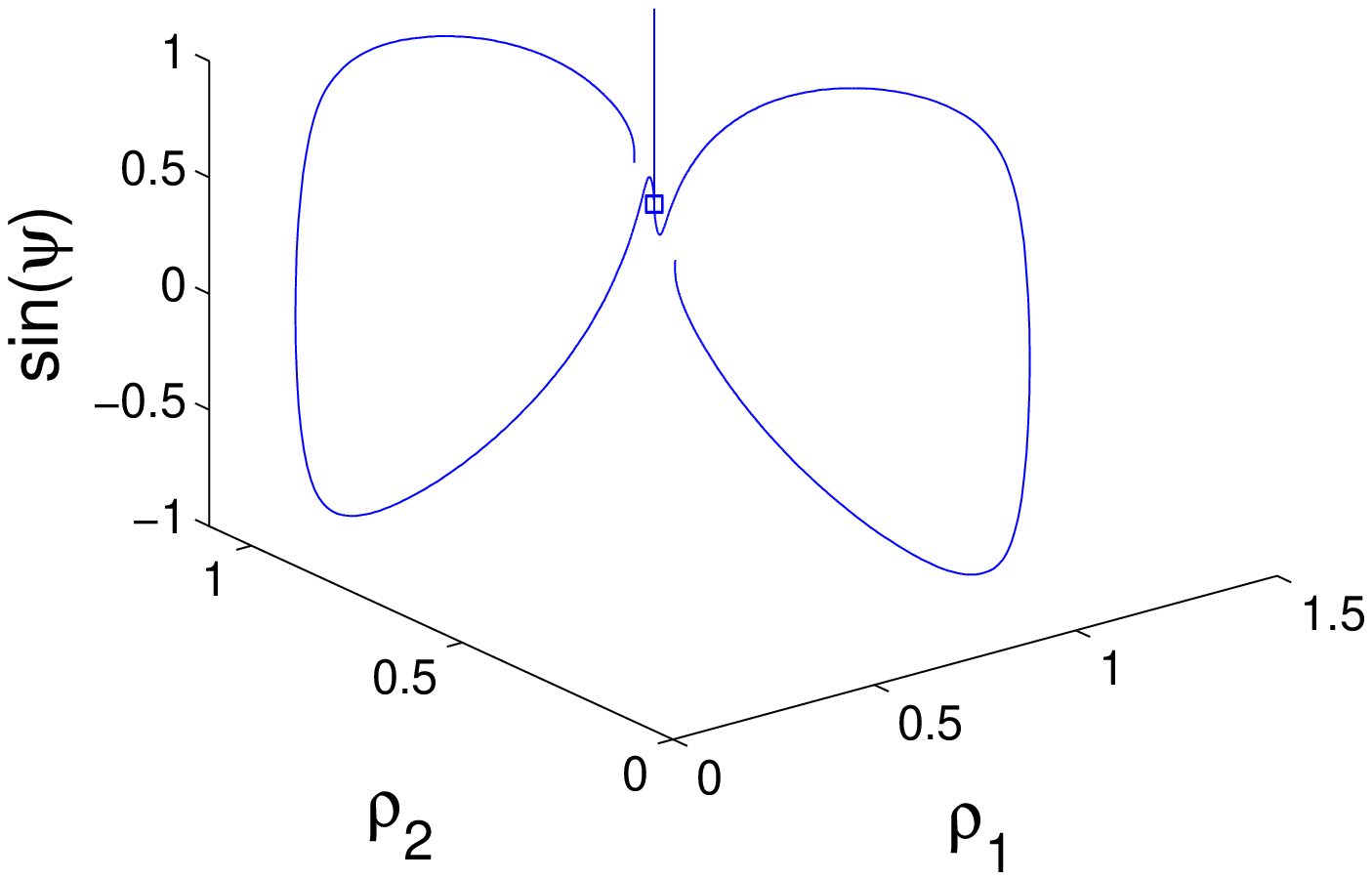}}\\
\subfigure[]{\includegraphics[width=0.32\linewidth]{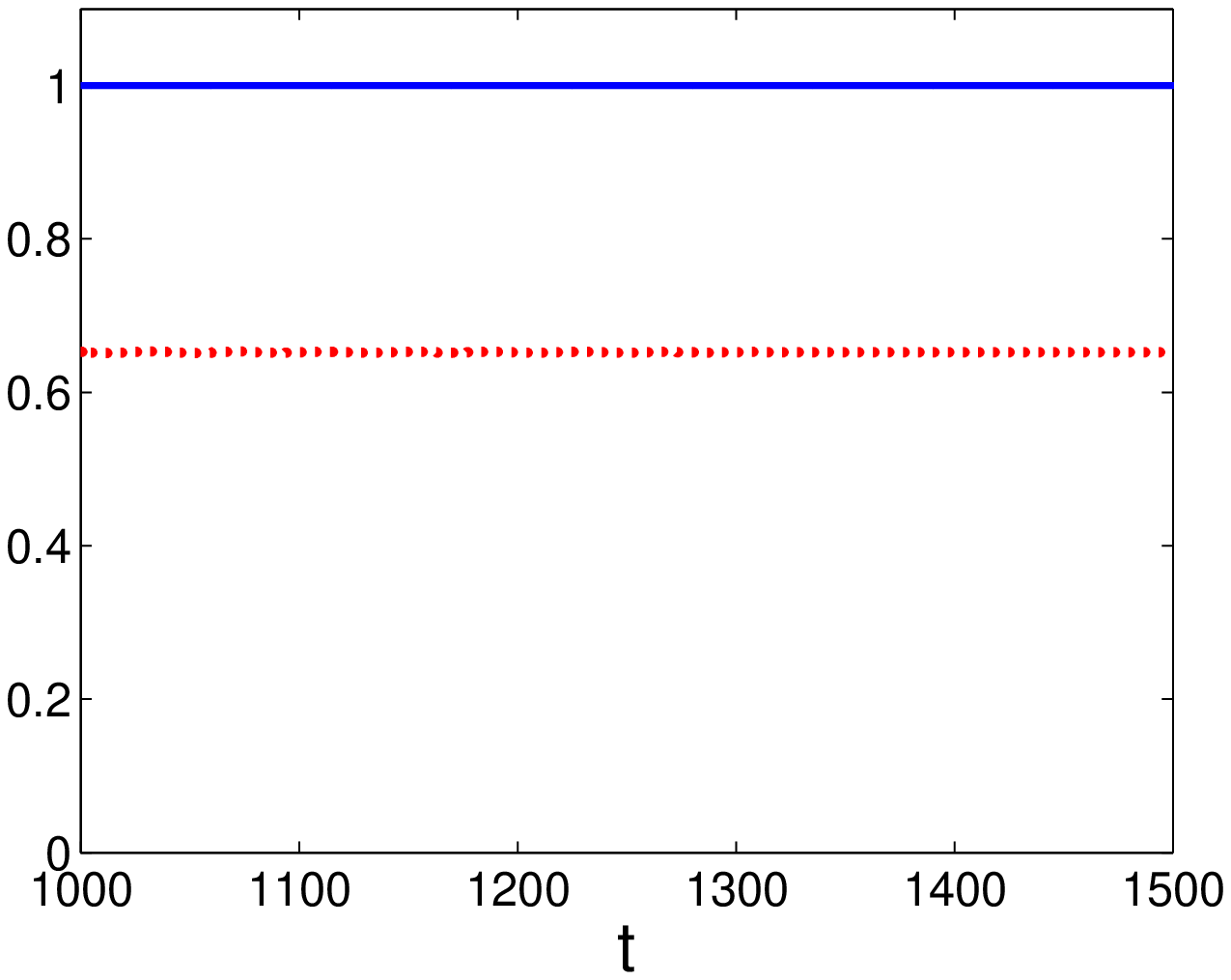}\label{fig:chimClassicoModello_b}}
\subfigure[]{\includegraphics[width=0.32\linewidth]{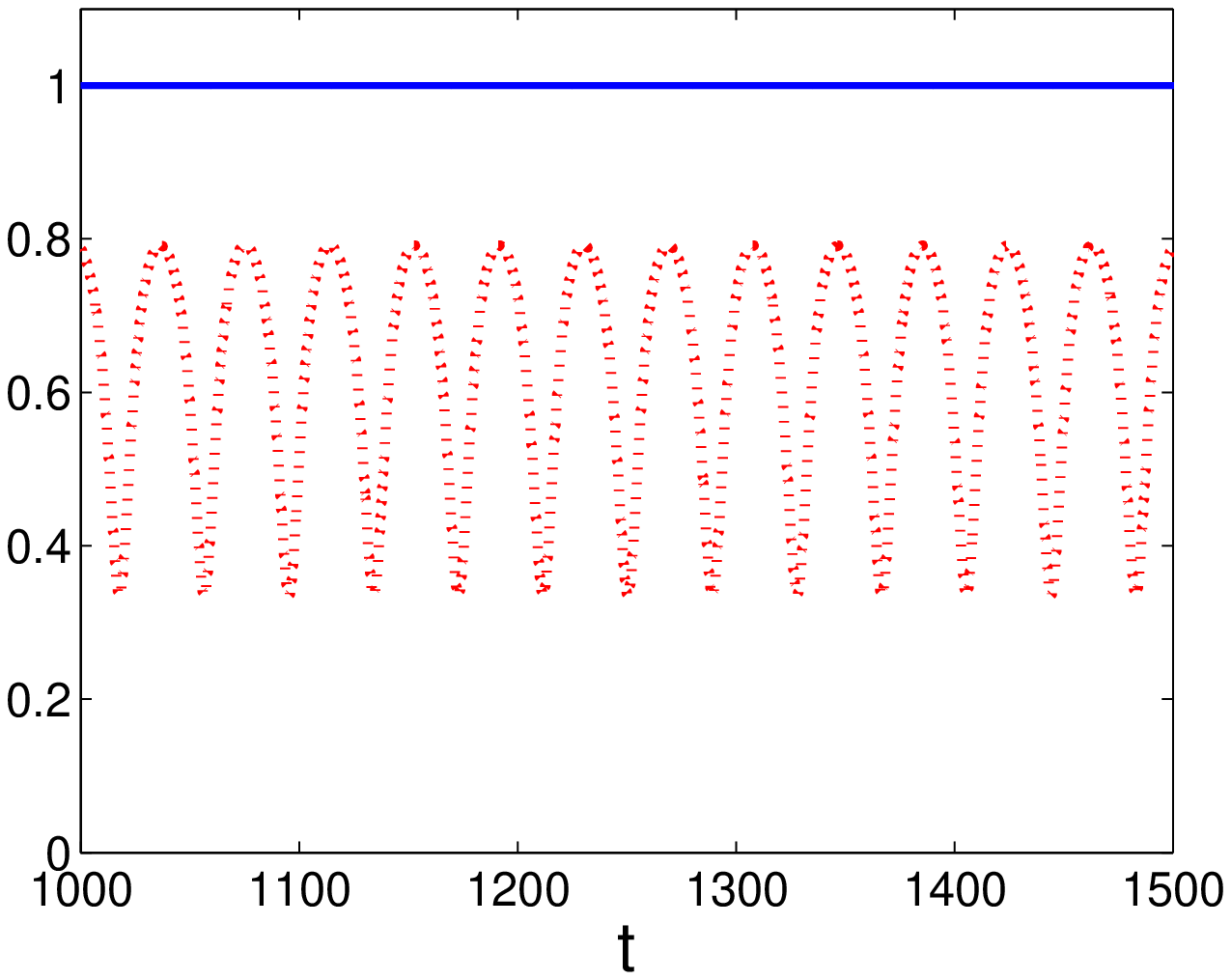}\label{fig:1b}}
\subfigure[]{\includegraphics[width=0.32\linewidth]{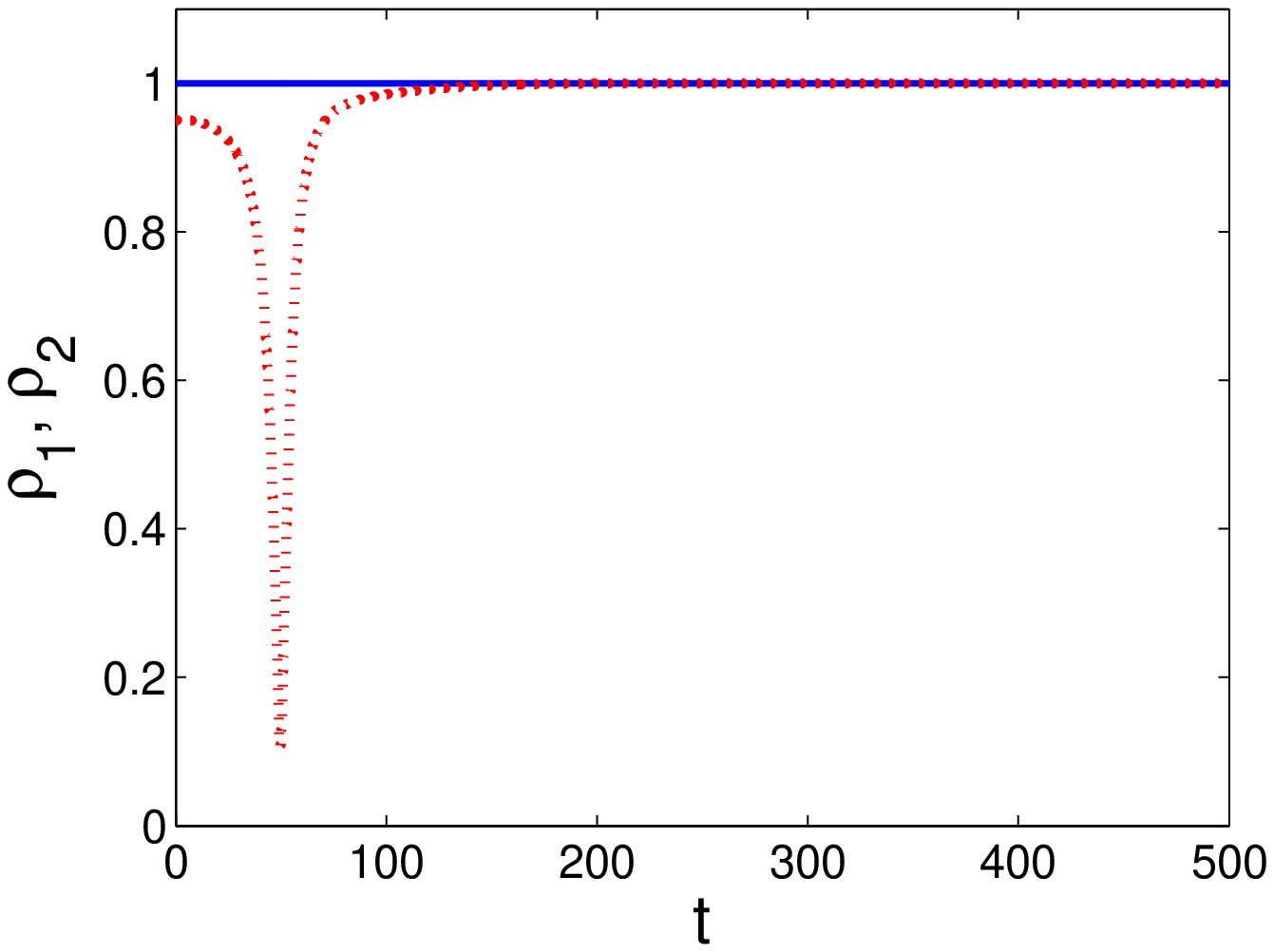}\label{fig:modello_p025}}
\caption{\label{fig:comparisonsenzarumore}(Color online)  Phase portrait of Eq.~(\ref{eq:rho1_rho2_psi2}) (a)-(c) and time evolution of $\rho_1$ (blue solid) and $\rho_2$ (red dotted) (d)-(f) with parameters: (a),(d) $p=0.38$; (b),(e) $p=0.33$; (c),(f) $p=0.25$. Other parameters as in Fig~\ref{fig:bif}.}
\end{figure*}

Considering again small switching periods $\tau$ and $p \in [C,0.5]$, we show now that the trajectory of the switching system~(\ref{eq:modelcoupledpopulations}) is close to that of the averaged reduced model~(\ref{eq:rho1_rho2_psi2}). For instance, for $p=0.38$ the phase-portrait and the trajectories of the averaged reduced model, shown in Fig.~\ref{fig:comparisonsenzarumore}(a),(d), are in agreement with the stable chimera displayed by the switching system in Fig.~\ref{fig:CSall}(a). An agreement is also found in the case of breathing chimera states, observed in the reduced model for $p \in [C,B]$. For example, for $p=0.33$ the state of the switching system (Fig.~\ref{fig:CSall}(b)) is a breathing chimera which is reproduced by the averaged reduced model (Fig.~\ref{fig:comparisonsenzarumore}(b),(e)).

\section{Onset of alternating chimeras}
\label{sec:alternatingchimera}
In this Section, we discuss the behavior in the region of parameter $p \in [0,C]$. In this region the reduced model predicts that only global synchronization is possible. However, in this region,
alternating chimera states are observed even for small switching intervals as shown in Fig.~\ref{fig:bifDiagram}. The discrepancy is due to the finite size of the network under consideration, which counts $N=100$ oscillators in each population.

\begin{figure}
\centering
\includegraphics[width=\linewidth]{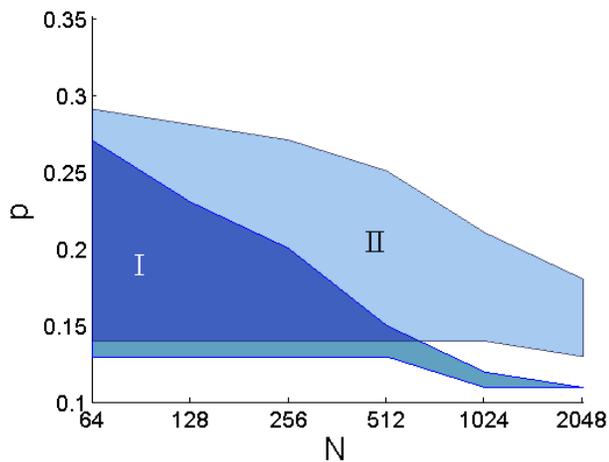}
\caption{(Color online) Extent of the region of alternating chimera as a function of the network size for two values of $\tau$. Region I corresponds to $\tau=0.1$, region II to $\tau=10$. Other parameters as in Fig.~\ref{fig:bifDiagram}. \label{fig:pMpm_N}}
\end{figure}

To show that the model accurately predicts the absence of chimera states in this region in the thermodynamic limit, we have carried out simulations at increasing values of the network size and identified the region where alternating chimera states appear. We notice that this region depends on $\tau$, but as predicted by the reduced model the one corresponding to small $\tau$ tends to shrink when the network size increases (Fig.~\ref{fig:pMpm_N}). For large $N$ alternating chimera states are still found for larger values of $\tau$ (for example, $\tau=10$ in Fig.~\ref{fig:pMpm_N}). We thus conjecture that the onset of chimera states is due to fluctuations and that the causes of these fluctuations are the finite size of the network and the large switching periods. In line with this, we show that: i) when noise is added to the reduced model, an alternating behavior may be observed; ii) fluctuations increase when $\tau$ is increased or $N$ decreased.

In the region $p \in [0,C]$, due to the homoclinic bifurcation, the structure of the phase portrait of the reduced model is such that the trajectory, starting in a neighborhood of the only stable equilibrium point $(1,1,0)$, experiences a large excursion before returning to the equilibrium (Fig.~\ref{fig:comparisonsenzarumore}(c),(f)). In presence of fluctuations, this may lead to a series of pulses in the evolution of the variables $\rho_1$ and $\rho_2$. To confirm this, simulations of the model~(\ref{eq:rho1_rho2_psi2}) subject to an additive noise term are carried out.
In particular, we have considered a stochastic term added to the averaged reduced system as follows
\begin{subequations}
\label{eq:rho1_rho2_psi2RUMORE}
\begin{align}
\dot{\rho}_1=& \frac{1-\rho_1^2}{2}[\mu \rho_1 \cos \alpha + p \rho_2 \cos(-\psi-\alpha)] + \xi(t)\\
\dot{\rho}_2=& \frac{1-\rho_2^2}{2}[\mu \rho_2 \cos \alpha + p \rho_1 \cos(\psi-\alpha)] - \xi(t)\\
\dot{\psi}=& -\frac{1+\rho_1^2}{2}\left[\mu \sin \alpha + p \frac{\rho_2}{\rho_1} \sin(\psi+\alpha)\right] \nonumber\\
 & + \frac{1+\rho_2^2}{2}\left[\mu \sin \alpha + p \frac{\rho_1}{\rho_2} \sin(-\psi+\alpha)\right],
\end{align}
\end{subequations}
where $\xi(t)$ is a Gaussian white noise satisfying $\langle \xi(t) \xi(t') \rangle = D \delta(t-t')$ with noise intensity $D$.

\begin{figure}
\centering
\subfigure[]{\includegraphics[width=0.48\linewidth]{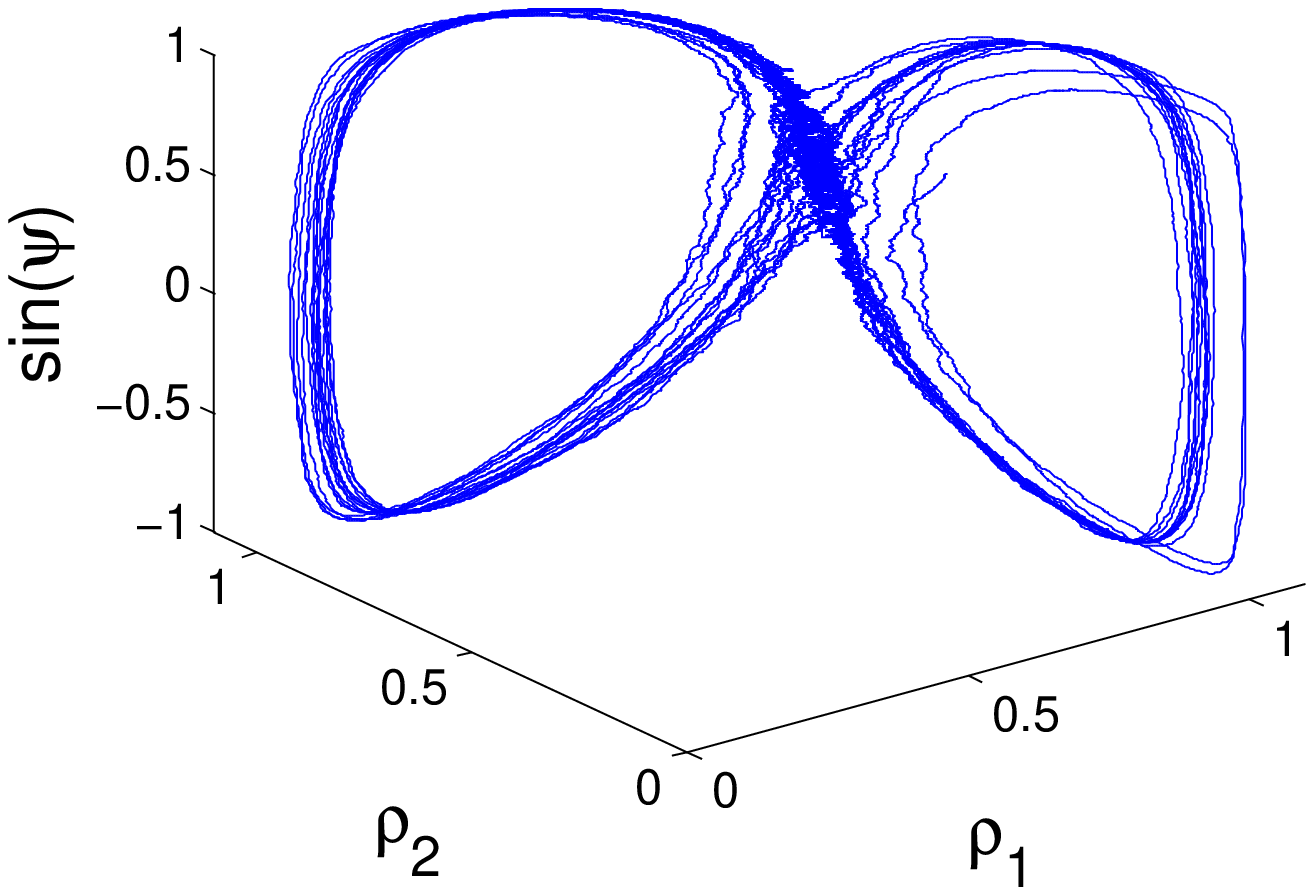}}
\subfigure[]{\includegraphics[width=0.48\linewidth]{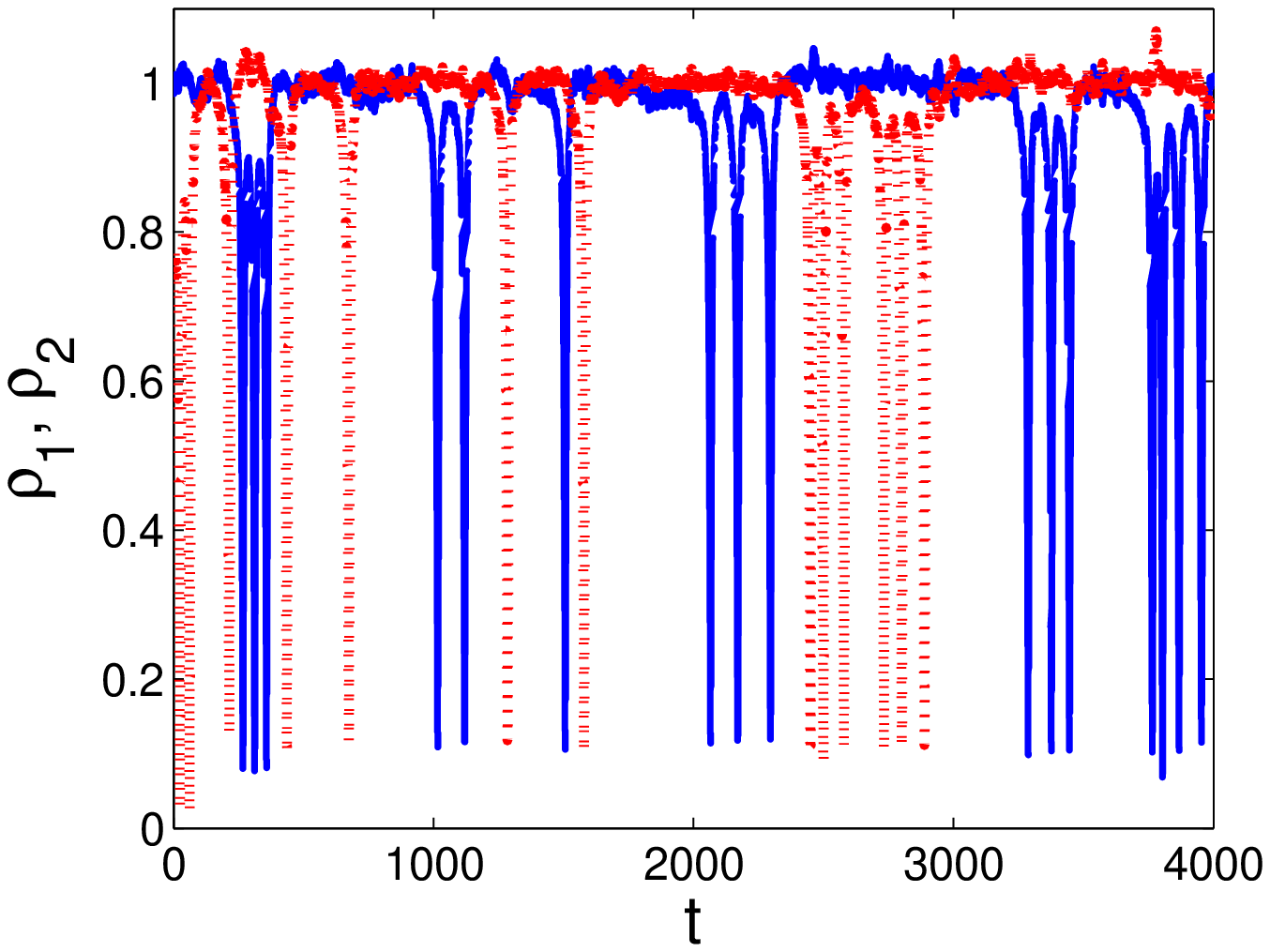}\label{fig:modello_p025noise}}
\caption{\label{fig:comparisonconrumore}(Color online)  Phase portrait of Eqs.~(\ref{eq:rho1_rho2_psi2RUMORE}) (a) and time evolution of $\rho_1$ (blue solid) and $\rho_2$ (red dotted) (b) with: $p=0.25$, $D=0.004$. Other parameters as in Fig.~\ref{fig:bifDiagram}.}
\end{figure}

We have numerically verified that a small level of noise in Eqs.~(\ref{eq:rho1_rho2_psi2RUMORE}) leads to switching chimera states analogous to those observed in the switching system. For instance, the alternating chimera state of Fig.~\ref{fig:CSall}(c) is also identified in the averaged reduced model (\ref{eq:rho1_rho2_psi2RUMORE}) for $p=0.25$ and $D=0.004$ (Fig.~\ref{fig:comparisonconrumore}).

\begin{figure}
\centering
\includegraphics[width=\linewidth]{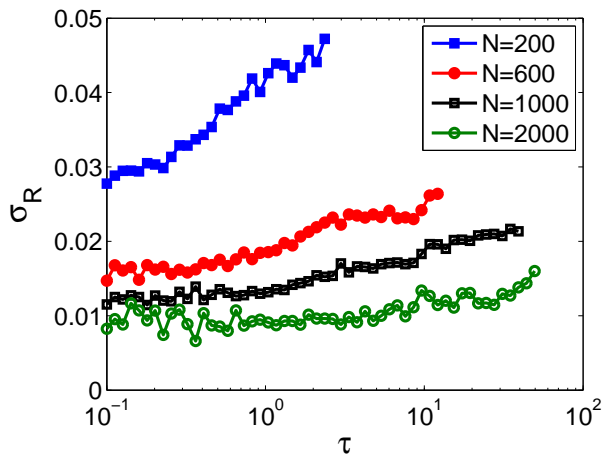}
\caption{Fluctuations as a function of $\tau$ for different values of $N$ measured by monitoring the standard deviation $\sigma_R$ of the Kuramoto order parameter of the desynchronized population of a stable chimera state for fixed switching probability $p=0.38$. Other parameters as in Fig.~\ref{fig:bifDiagram}.
\label{fig:std_tauN1}}
\end{figure}

To show the dependence of the fluctuations on $\tau$ and $N$, we have measured the standard deviation $\sigma_R$ of the Kuramoto order parameter in Eq.~(\ref{eq:order_param}) for the desynchronized population of a stable chimera state  in dependence on $\tau$ at different values of $N$ and for fixed $p=0.38$. Each curve corresponding to a different $N$ is calculated while the value of $\tau$ is such that a stable chimera state is obtained and stopped when the behavior bifurcates to global synchronization. The parameter is reported in Fig.~\ref{fig:std_tauN1} which shows how indeed fluctuations of the order parameter increase with $\tau$ and decrease with $N$.

% \textbf{Is it possible to measure the fluctuations of
% some parameter of the time-varying network? Maybe this could have a better statistics.}

% \textbf{We have observed that alternating chimeras are also found in a system of three coupled populations. Do we want to include a small paragraph on this?}

\section{Conclusions}
\label{sec:conclusion}
In this work, we have considered a pair of two populations of identical oscillators with time-varying inter-population links. In the case of fixed connectivity, such a network exhibits stable or breathing chimeras, while alternating chimeras may be observed only if a degree of heterogeneity in the distribution of oscillator intrinsic frequencies is introduced. When the inter-population links change over time, we have found that the network may support all the three chimera states even in the case of identical oscillators.

The switching between the different network topologies, which result from the stochastic rule used to establish inter-population links, induces fluctuations in the system. We have found that such fluctuations are averaged out in the thermodynamic limit and under the assumption of small switching intervals. In this case, the dynamics of the system can be qualitatively represented by a low-dimensional averaged system that accurately predicts the stable and breathing chimeras. However, fluctuations are fundamental to explain the onset of alternating chimera states and can be incorporated in the low-dimensional model with the addition of a stochastic term. Since fluctuations increase for decreasing values of $N$ and increasing of $\tau$, alternating chimera states are likely to occur not only in small networks, but also in arbitrary large structures in the presence of large switching time intervals.

Our findings can be generalized for more than two populations coupled in a ring configuration with time-varying inter-population links. We have evidence that this gives rise to traveling incoherent domains and other spatio-temporal patterns of coherence and incoherence.

\begin{acknowledgments}
The authors would like to thank Thomas Isele for helpful discussions. PH acknowledges supported by Deutsche Forschungsgemeinschaft in the framework of SFB 910.
\end{acknowledgments}

\end{document}